\documentclass[aps,pre,12pt]{revtex4}

\usepackage{graphicx}
\usepackage{subfigure}

\begin{document}

\title{The structure of weak shocks in quantum plasmas}

\author{Vitaly Bychkov, Mikhail Modestov and Mattias Marklund  }

\affiliation{Department of Physics, Ume\aa\ University, SE--901 87
Ume\aa, Sweden}

\begin{abstract}
The structure of a weak shock in a quantum plasma is studied, taking into account both dissipation terms due to thermal conduction and dispersive quantum terms due to the Bohm potential. Unlike quantum systems without dissipations, even a small thermal conduction may lead to a stationary shock structure. In the limit of zero quantum effects, the monotonic Burgers solution for the weak shock is recovered. Still, even small quantum terms make the structure non-monotonic with the shock driving a train of oscillations into the initial plasma. The oscillations propagate together with the shock. The oscillations become stronger as the role of Bohm potential increases in comparison with thermal conduction. The results could be of importance for laser-plasma interactions, such as inertial confinement fusion plasmas, and in astrophysical environments, as well as in condensed matter systems.
\end{abstract}

\maketitle

\section{Introduction}

Quantum plasmas, where the finite width of the electron wave functions gives rise to collective effects \cite{Pines,Pines-book,Kremp-etal}, are currently a rapidly growing field of research. Many of the studies are motivated by the potential for application to nanoscale systems \cite{craighead}, such as quantum wells \cite{manfredi-hervieux}, ultracold plasmas \cite{robinson-etal,fletcher-etal}, laser fusion plasmas \cite{glenzer-etal}, next-generation high intensity light sources \cite{Marklund-Shukla,Mourou-etal}, and plasmonic devices \cite{SPP}. Moreover, nonlinear effects in quantum plasmas, such as the formation of dark solitons and vortices \cite{shukla-eliasson}, interaction between quantum plasma oscillations and electromagnetic waves \cite{shukla-eliasson2}, quantum turbulence \cite{shaikh-shukla}, and solitary structures \cite{brodin-marklund2,marklund-etal} supported by the electron spin \cite{marklund-brodin,brodin-marklund}, are currently in focus as well.

There has also been much interest in shocks in quantum-like systems, such as nonlinear optical fibers and Bose-Einstein condensates
\cite{Wan-Nature-2007,Kamchatnov.et.al-2002,Kamchatnov.et.al-2004,Damski-2004}. The structure of such quantum shocks is markedly different from the classical ones \cite{Landau-Lifshitz-Fluid}. The shock structure in classical fluids/gases is governed by transport processes, \textit{i.e.}, the viscosity and thermal conduction. A classical shock propagating with constant velocity displays a stationary structure. If the shock is weak, then transition from initial matter to  compressed one may be described by the smooth monotonic Burgers solution \cite{Landau-Lifshitz-Fluid}. In contrast to classical fluids, quantum media typically exhibit dispersion due to the Bohm potential instead of dissipation
\cite{Wan-Nature-2007,Kamchatnov.et.al-2002,Kamchatnov.et.al-2004,Damski-2004,Haas-et.al.-2003,Gardner-1994,Ali-et.al-2007,Misra-et.al-2007}.  For this reason, even a quantum shock propagating with constant velocity in a uniform medium does not posses a stationary structure. Transition from initial to compressed quantum media happens in the form of a train of solitons of different amplitudes
\cite{Wan-Nature-2007,Kamchatnov.et.al-2002,Kamchatnov.et.al-2004,Damski-2004}. The solitons propagate with different velocities, which makes the whole structure intrinsically non-stationary. Obviously, a train of solitons also provides a non-monotonic transition from initial to final state of the medium. However, there are quantum systems with both dissipations and dispersion, such as quantum plasmas. The viscosity in plasma is determined by ions and it is typically negligible. Still, electron thermal conduction may be quite strong both in classical and quantum plasmas \cite{Gardner-1994}. Therefore, shocks in such plasmas may demonstrate transitional behavior between the classical and quantum domains. The purpose of the present paper is to trace such a transition by studying weak shocks.

Here, we derive a nonlinear equation governing the structure of a weak shock in quantum plasma. The equation contains both dissipation terms (due to thermal conduction) and dispersive quantum terms (due to the Bohm potential). Unlike quantum systems without dissipation, even relatively weak thermal conduction may lead to a stationary structure of a shock. In the limit of zero quantum effects we recover the monotonic Burgers solution for the shock structure. Still, even small quantum terms make the transition non-monotonic with the shock driving a train of oscillations into the initial plasma. The oscillations propagate together with the shock with the same velocity. The oscillations become stronger as the role of Bohm potential increases in comparison with thermal conduction. The oscillations resemble the soliton train in quantum shocks without dissipations.

\section{Governing equations}
	
The basic equation of nonrelativistic quantum mechanics is the Schr\"odinger
equation. The dynamics of an electron, represented by its wave function $\psi$, in an external electromagnetic field
$(\phi,\mathbf{A})$ is governed by 
\begin{equation}\label{eq:schrodinger}
  i\hbar\frac{\partial \psi}{\partial t} + \frac{\hbar}{2 m_e}\left(\nabla + \frac{ie}{\hbar}\mathbf{A} \right)^2\psi + e\phi\psi  = 0 ,
\end{equation}
where $\hbar$ is Planck's constant, $m_e$ is the electron mass, and $e$ is the magnitude
of the electron charge. This complex equation may be written as two real equations, writing 
$\psi = \sqrt{\rho}\,\exp{iS/\hbar}$, where $\rho$ is the amplitude and $S$ the phase of the wave function, respectively \cite{holland}. Such a decomposition was presented by de Broglie and Bohm in order to 
understand the dynamics of the electron wave packet in terms of classical variables. In Ref.\ \cite{Gardner-1994} the Wigner function was employed for the purpose of obtaining a set of quantum hydrodynamic equations. In this way, an arbitrary number of conservation equations, in particular an energy conservation equation, may be obtained before closure. Here we will just briefly review the Bohm--de Broglie approach, making use of the energy conservation equation from Ref.\ \cite{Gardner-1994}.
Using the decomposition of the wave function in terms of its amplitude and phase, Eq.\ (\ref{eq:schrodinger}) gives
\begin{equation}\label{eq:schrod-cont}
  \frac{\partial \rho}{\partial t} + \nabla\cdot(\rho\mathbf{u}) = 0 , 
\end{equation}
and 
\begin{equation}\label{eq:schrod-mom}
  m_e\frac{d\mathbf{u}}{d t} = e(\mathbf{E} + \mathbf{u}\times\mathbf{B})
    + \frac{\hbar^2}{2m_e}\nabla\left(\frac{\nabla^2\sqrt{\rho}}{\sqrt{\rho}}\right) ,
\end{equation}
where the velocity is defined by $\mathbf{u} = \nabla S/m_e$, and $\mathbf{E} = -\nabla\phi - \partial_t\mathbf{A}$ and $\mathbf{B} = \nabla\times\mathbf{A}$. The last term of Eq.\ (\ref{eq:schrod-mom})
is the gradient of the Bohm--de Broglie potential, and is due to the effect of wave function dispersion. We also note the
striking resemblance of Eqs.\ (\ref{eq:schrod-cont}) and (\ref{eq:schrod-mom}) to the classical fluid equations.

Suppose that we have $N$ electron wavefunctions, independent apart from their interaction via the electromagnetic field. For each wave function $\psi_\alpha$, we have a corresponding probability
$\mathcal{P}_\alpha$. From this, we first define $\psi_\alpha = \sqrt{\rho_\alpha}\exp(iS_\alpha/\hbar)$ and follow the steps leading to Eqs. (\ref{eq:schrod-cont}) and (\ref{eq:schrod-mom}). We now have $N$ such equations the wave functions $\{\psi_\alpha\}$. Defining
\begin{equation}\label{eq:meandensity}
  \rho \equiv \sum_{\alpha = 1}^N \mathcal{P}_\alpha \rho_\alpha 
\end{equation}
and 
\begin{equation}\label{eq:meanvelocity}
  \mathbf{u} \equiv \langle \mathbf{u}_\alpha \rangle = \sum_{\alpha = 1}^N \frac{\mathcal{P}_\alpha \rho_\alpha \mathbf{u}_\alpha}{\rho} ,
\end{equation}
we can define the deviation from the mean flow according to 
\begin{equation}
  \mathbf{w}_\alpha = \mathbf{u}_\alpha - \mathbf{u} .
\end{equation}
Taking the average, as defined by (\ref{eq:meanvelocity}), of Eqs. (\ref{eq:schrod-cont}) and (\ref{eq:schrod-mom}) and using the above variables, we obtain the quantum fluid equation
\begin{equation}\label{eq:cont}
  \frac{\partial \rho_e}{\partial t} + \nabla\cdot(\rho\mathbf{u}) = 0  
\end{equation}
and 
\begin{equation}\label{eq:mom-schrod}
  \rho_e\left(\frac{\partial}{\partial t} + \mathbf{u}_e\cdot\nabla\right)\mathbf{u}_e = \frac{e\rho_e}{m_e}\left( \mathbf{E} + \mathbf{u}_e\times\mathbf{B}\right) - \nabla P_e + \frac{\hbar^2\rho_e}{2m_e^2}\nabla\left\langle \left( \frac{\nabla^2\sqrt{\rho_\alpha}}{\sqrt{\rho_\alpha}}\right) \right\rangle ,
\end{equation}
where we have assumed that the average produces an isotropic pressure $P = \rho_e\langle |\mathbf{w}_\alpha|^2\rangle$
We note that the above equations still contain an explicit sum over the electron wave functions. For typical scale lengths larger than the Fermi wavelength $\lambda_F $, we may approximate the last term by the Bohm--de Broglie potential \cite{Gardner-1994}
\begin{equation}
  \left\langle \frac{\nabla^2\sqrt{\rho_\alpha}}{\sqrt{\rho_\alpha}} \right\rangle \approx  
  \frac{\nabla^2\sqrt{\rho_e}}{3\sqrt{\rho_e}} ,
\end{equation}
where the factor $1/3$ comes from an isotropic averaging. 

Treating the ions as fully classical due to their large inertia, we can derive a a set of single-fluid quantum magnetohydrodynamic (MHD) equations, following \cite{brodin-marklund} where a similar set of equations
for a perfect fluid type quantum MHD system was derived. 
Thus, we obtain the equations of mass, momentum and energy transfer in a quantum plasma \cite{Gardner-1994,brodin-marklund,Haas}
\begin{equation}\label{eq.1}
  \frac{\partial \rho}{\partial t} + \frac{\partial}{\partial x_{j}}(\rho u_{j}) = 0 ,
\end{equation}
\begin{eqnarray}
  \frac{\partial }{\partial t} (\rho u_{j}) + \frac{\partial}{\partial x_{l}}(\rho u_{j} u_{l})
    =-\frac{\partial P}{\partial x_{j}}
  +
   \frac{\hbar^2}{12m_em_i}
\frac{\partial}{\partial x_{l}}
    \left(\rho\frac {\partial^{2} }{\partial x_j\partial x_l}\ln{\rho}\right) ,
  \label{eq.2}
  \end{eqnarray}
and
  \begin{eqnarray}
  && \frac{\partial }{\partial t} \left[\rho \varepsilon + \rho \frac{u^{2}}{2}  - \frac{\hbar^2}{24m_em_i}\rho \nabla^{2} \ln{\rho}\right]
  + \frac{\partial}{\partial x_{j}} \Bigg[\rho u_{j}
  \left(h + \frac{u^{2}}{2}  - \frac{\hbar^2}{24m_em_i} \nabla^{2} \ln{\rho}\right)
  \nonumber\\ && \qquad\qquad
  - \rho u_{l}
   \frac{\hbar^2}{12m_em_i} \frac {\partial^{2} \ln{\rho}}{\partial x_j\partial x_l}  - \kappa \frac{\partial T}{\partial x_{j}}\Bigg] = 0,
  \label{eq.3}
  \end{eqnarray}
where $\rho$  is the fluid mass density, $u_j$ is the fluid velocity, $P$ is the pressure, $\varepsilon$, $h$  are thermal energy and enthalpy, $\kappa$  is thermal conduction, and $m_i$, $m_e$  are the ion and electron masses. The energy conservation equation (\ref{eq.3}) is derived using the Wigner approach \cite{Gardner-1994} for the electron dynamics and combining this with the ion equation in the MHD limit. Here we have neglected the effects due to the magnetic field, that are assumed small in comparison to the other governing terms. Such terms can easily be included \cite{Haas,brodin-marklund}. The hydrodynamic equations should be complemented by the thermodynamic equation of state. We take the equation of state to be that of an ideal gas, \textit{i.e.},
\begin{equation}\label{eq.4}
  P=\frac{\gamma - 1}{\gamma} C_{P} \rho T ,
\end{equation}
and
\begin{equation}\label{eq.5}
  h=C_{P} T,
\end{equation}
and let the electron thermal conduction $\kappa$ be $\propto T^{5/2}$. Here $C_{P}$  is heat capacity at constant pressure and  $\gamma$ is the adiabatic exponent. We stress that the forms (\ref{eq.4}) and (\ref{eq.5}) play a minor role for weak shocks.

\section{Shock solutions}

We consider a planar stationary shock. In the reference frame of the shock, Eqs. (\ref{eq.1})--(\ref{eq.3}) may be integrated as
\begin{equation}\label{eq.7}
  \rho u = \rho_{0} u_{0},
\end{equation}
\begin{equation}\label{eq.8}
  P + \rho u^{2} - \frac{\hbar^{2}}{12 m_{e}m_{i}} \rho \frac{d^{2} \ln \rho }{dx^{2}}= P_{0}+\rho_{0} u_{0}^{2},
\end{equation}
and
\begin{equation}\label{eq.9}
  h +  \frac{u^{2}}{2} - \frac{\hbar^{2}}{8 m_{e}m_{i}} \frac{d^{2} \ln \rho }{dx^{2}} - \frac{\kappa}{\rho_{0}u_{0}} \frac{dT}{dx}= h_{0}+ \frac{u_{0}^{2}}{2},
\end{equation}
where the subscript 0 refers to the uniform plasma ahead of the shock and $u_{0}$ is the shock speed. As we can see from (\ref{eq.7})--(\ref{eq.9}), quantum effects do not influence the properties of the uniform flow behind the shock; they are important only for the shock structure. Next, we introduce the parameters
\begin{equation}\label{eq.13}
  L= \frac{\kappa_{0}}{C_{P}\rho_{0}u_{0}},
\end{equation}
and
\begin{equation}\label{eq.14}
  \mathrm{Ma}^{2}=\frac{\rho_{0} u_{0}^{2}}{\gamma P_{0}},
\end{equation}
and the scaled variables  $\rho/\rho_{0}=u_{0}/u=R$,  $T/T_{0}=1+\vartheta$,  $\eta =x/L$. Here   $L$ is the characteristic length scale determined by thermal conduction; in the classical case   it may be treated as the shock width with the accuracy of a numerical factor of order unity. The other parameter is the Mach number, $\mathrm{Ma}$, which compares the shock velocity $u_{0}$  to the initial sound speed $\sqrt{\gamma P_{0}/\rho_{0}}$ in the plasma and characterizes the shock strength. The parameter $\vartheta$ denotes the deviation of the temperature, produced by the shock wave, from the initial value. In the case of weak shocks we have $\mathrm{Ma} - 1 \ll 1$, \textit{i.e.}, the shock velocity marginally exceeds the sound speed, and as does the temperature from the initial value $\vartheta \ll 1$. Using the scaled variables, we reduce Eqs. (\ref{eq.7})--(\ref{eq.9}) to
\begin{equation}\label{eq.23}
  R(1+\vartheta) + \frac{\gamma \mathrm{Ma}^{2}}{R} - Q R  \frac{d^{2} \ln R }{d\eta^{2}}= 1+\gamma \mathrm{Ma}^{2},
\end{equation}
and
\begin{eqnarray}
  \vartheta + \frac{\gamma -1 }{2R^{2}} \mathrm{Ma}^{2}- (1+\vartheta)^{5/2}\frac{d\vartheta}{d\eta}
  -
   \frac{3(\gamma - 1)}{2\gamma}Q \frac{d^{2} \ln R }{d\eta^{2}}= \frac{\gamma-1}{2} \mathrm{Ma}^{2} ,
  \label{eq.24}
  \end{eqnarray}
where
\begin{equation}\label{eq.25}
  Q=\frac{\hbar^{2}\rho_{0}L^{2}}{12m_{e}m_{i}P_{0}}
\end{equation}
is the parameter comparing the role of quantum and classical effects in the shock dynamics. This parameter can be interpreted as a quantum Mach number. The system (\ref{eq.23})--(\ref{eq.24}) determines the structure of a shock wave in a quantum plasma. In the classical case we have $Q=0$, and Eq.\ (\ref{eq.23})  gives an algebraic relation between the density and temperature
\begin{equation}\label{eq.23a}
  \frac{1}{R}=\frac{1+\gamma \mathrm{Ma}^{2}}{2\gamma \mathrm{Ma}^{2}}\left[1\pm \sqrt{1-\frac{4\gamma \mathrm{Ma}^{2}(1+\vartheta)}{(1+\gamma \mathrm{Ma}^{2})^{2}}}\right].
\end{equation}
The positive sign in (\ref{eq.23a}) gives rise to shock solutions, while the negative sign corresponds to deflagrations \cite{Landau-Lifshitz-Fluid}. We note that the density and temperature of the compressed matter increase together in a shock. In deflagrations, the temperature increase leads to a decrease of the density, as in, \textit{e.g.}, laser ablation and flames
\cite{Manheimer.et.al-1982,Bychkov.et.al-1994,Betti.et.al-1996,Bychkov-Liberman-2000}. Substituting (\ref{eq.23a}) into (\ref{eq.24}), we obtain a single differential equation for the temperature in a classical shock. In the case of strong quantum shocks, one has to solve a system of two differential equations.

\subsection{Weak shocks}

In the present paper we investigate only the case of a weak shock  with $\mathrm{Ma}^{2}-1=\mu \ll 1$  and  $\vartheta \ll 1$. A more general case will be studied elsewhere. As note above, this value of $\mu$ characterizes a shock velocity marginally above the sound speed. In the case of a weak shock in the linear approximation, Eq. (\ref{eq.23a}) may be simplified according to
\begin{equation}\label{eq.26a}
  \frac{1}{R}=1-\frac{\vartheta}{\gamma \mathrm{Ma}^{2}-1}.
\end{equation}
Taking into account the quantum dispersion and the weak nonlinearity in (\ref{eq.23}), we find
\begin{eqnarray}
  \frac{1}{R}=1-\frac{\vartheta}{\gamma \mathrm{Ma}^{2}-1}
  +
   \frac{Q}{\gamma \mathrm{Ma}^{2}-1}\frac{d^{2}\ln R}{d\eta^{2}}-\frac{\gamma \mathrm{Ma}^{2}\vartheta^{2}}{(\gamma \mathrm{Ma}^{2}-1)^{3}} .
  \label{eq.26}
  \end{eqnarray}
Equation (\ref{eq.26}) relates the density to the temperature in a weak shock. Taking into account the linear approximation (\ref{eq.26a}), we may simplify the quantum dispersive term as
\begin{equation}\label{eq.26b}
 \frac{d^{2}\ln R}{d\eta^{2}} =\frac{1}{\gamma \mathrm{Ma}^{2}-1}\frac{d^{2}\vartheta}{d\eta^{2}}.
\end{equation}
Substituting (\ref{eq.26}) into (\ref{eq.24}), we find
\begin{equation}\label{eq.27}
 \vartheta \mu  - \left(\frac{\gamma + 1}{\gamma - 1}\right) \frac{\vartheta^{2}}{2} + \frac{3-\gamma}{2\gamma }Q\frac{d^{2}\vartheta}{d\eta^{2}}=(\gamma - 1) \frac{d\vartheta}{d\eta}.
\end{equation}
Equation (\ref{eq.27}) describes the structure of a weak shock in a quantum plasma with finite thermal conduction. Dissipation and quantum terms enter Eq.\ (\ref{eq.27}) as the first and second derivatives of the scaled temperature. All derivatives tend to zero in the uniform flows corresponding to the initial and final plasma states at  $\eta \rightarrow \pm \infty$. Taking into account that $\vartheta = 0$  in the initial flow, we find the relation between scaled temperature increase in the shock wave and the scaled shock speed
\begin{equation}\label{eq.20a}
 \mu  = \left(\frac{\gamma + 1}{\gamma - 1}\right) \frac{\vartheta_{1}}{2} .
\end{equation}
Thus, we can rewrite (\ref{eq.27}) as
\begin{equation}\label{eq.28}
 \vartheta (\vartheta_{1} - \vartheta) = \frac{2(\gamma - 1)^{2}}{\gamma + 1} \frac{d\vartheta}{d\eta} - \frac{(3-\gamma)(\gamma - 1)}{\gamma (\gamma + 1) }Q\frac{d^{2}\vartheta}{d\eta^{2}}.
\end{equation}
The  solution to (\ref{eq.28})  changes from $\vartheta = 0$  at $\eta \rightarrow - \infty$  in the initial plasma ahead of the shock to $\vartheta = \vartheta_{1}$   at  $\eta \rightarrow  \infty$ in the compressed plasma behind the shock.

\subsection{Classical/quantum transition in the schock}

We are interested in solution to (\ref{eq.28}) for any parameter value $Q$ from $0$ (classical plasma) to infinity (quantum plasma without dissipations). To simplify our study of Eq. (\ref{eq.28}), we may rescale the  temperature according to $\phi = \vartheta / \vartheta_{1}$,  so that $\phi$  changes from $0$ in the initial plasma to $1$ in the compressed plasma. We also rescale the coordinate
\begin{equation}\label{eq.29a}
 \xi = \eta \frac{\vartheta_{1}(\gamma + 1)}{2(\gamma - 1)^{2}} ,
\end{equation}
and introduce a new parameter
\begin{equation}\label{eq.29b}
 q = Q \vartheta_{1} \frac{(\gamma + 1)(3-\gamma)}{2\gamma(\gamma - 1)^{3}} ,
\end{equation}
Then (\ref{eq.28}) reduces to a concise form
\begin{equation}\label{eq.29}
 -q\frac{d^{2}\phi}{d\xi^{2}} + \frac{d\phi}{d\xi} = \phi(1-\phi).
\end{equation}
The parameter $q$  describes the relative role of quantum effects and thermal conduction in the shock.
Equation (\ref{eq.29}) is the main result of our paper.

In the case of zero quantum effects ($q=0$), Eq. (\ref{eq.29}) goes over to the stationary Burgers equation with the solution
\begin{equation}\label{eq.30}
 \phi=\frac{\exp \xi}{1+\exp \xi}.
\end{equation}
The influence of quantum effects may be analyzed analytically in the limit of small  $q \ll 1$. Because of the invariance to a shift in space, the solution to (\ref{eq.29}) may be presented the form
\begin{equation}\label{eq.31}
 \frac{d\phi}{d\xi}=f(\phi),\quad\quad
 \frac{d^{2}\phi}{d\xi^{2}}=f\frac{df}{d\phi},
\end{equation}
so that
\begin{equation}\label{eq.31a}
 \phi(1-\phi)=f\left(1-q\frac{df}{d\phi}\right).
\end{equation}
To $0^{\mathrm{th}}$ order in $q$ $(\ll 1)$ we have $
f(\phi)=\phi(1-\phi)$, and to $1^{\mathrm{st}}$ order in $q$, Eq. (\ref{eq.29}) becomes
\begin{equation}\label{eq.32}
 \frac{\phi(1-\phi)}{1-q(1-2\phi)}=\frac{d\phi}{d\xi}.
\end{equation}
The classical solution (\ref{eq.30}) was symmetric in space with respect to the central point $\phi = 1/2$. As we can see from (\ref{eq.32}), quantum effects make the scaled temperature slope $d\phi/d\xi$  steeper in the front part of the shock, for  $\phi < 1/2$, and smoother at the back side, for  $\phi > 1/2$.

\begin{figure}
\subfigure[]{\includegraphics[width=.8\columnwidth]{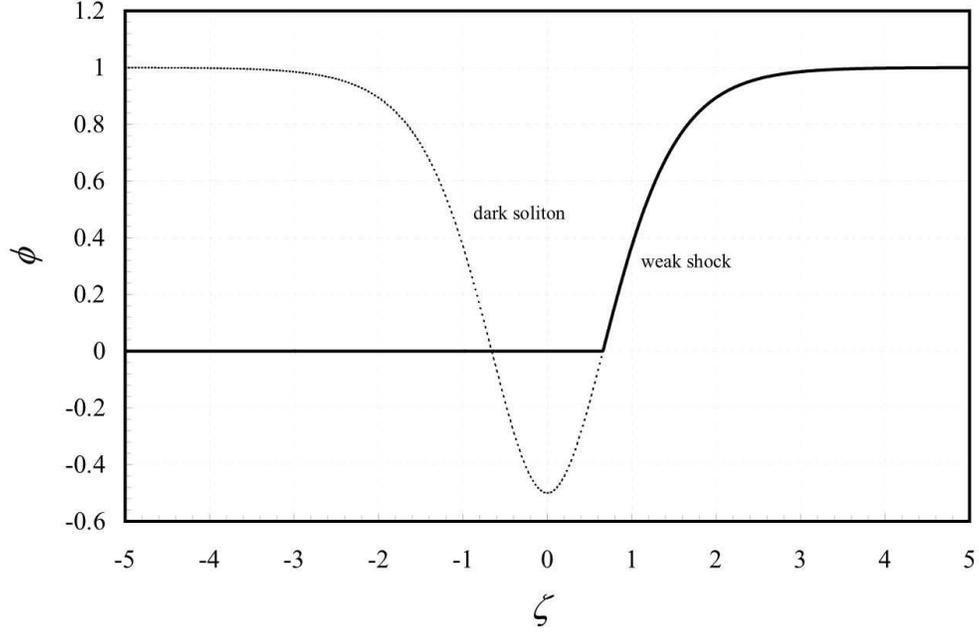}}
\subfigure[]{\includegraphics[width=.8\columnwidth]{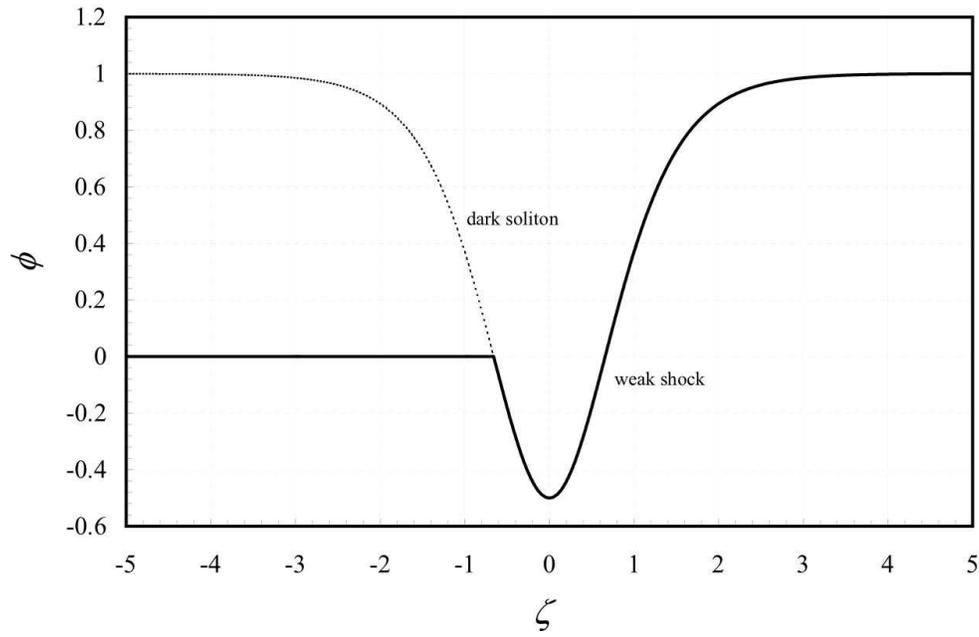}}
\caption{Scaled temperature $\phi$
versus scaled coordinate $\zeta = \xi /\sqrt{q}$ in the case of zero dissipations $q\rightarrow \infty$.
In the plots (a) and (b), temperature approaches the weak discontinuity at $\phi = 0$ from the "bright" and "dark" sides, respectively. The dashed line shows the soliton (\ref{eq.35}).} \label{fig-1}
\end{figure}

\begin{figure}
\includegraphics[width=.8\columnwidth]{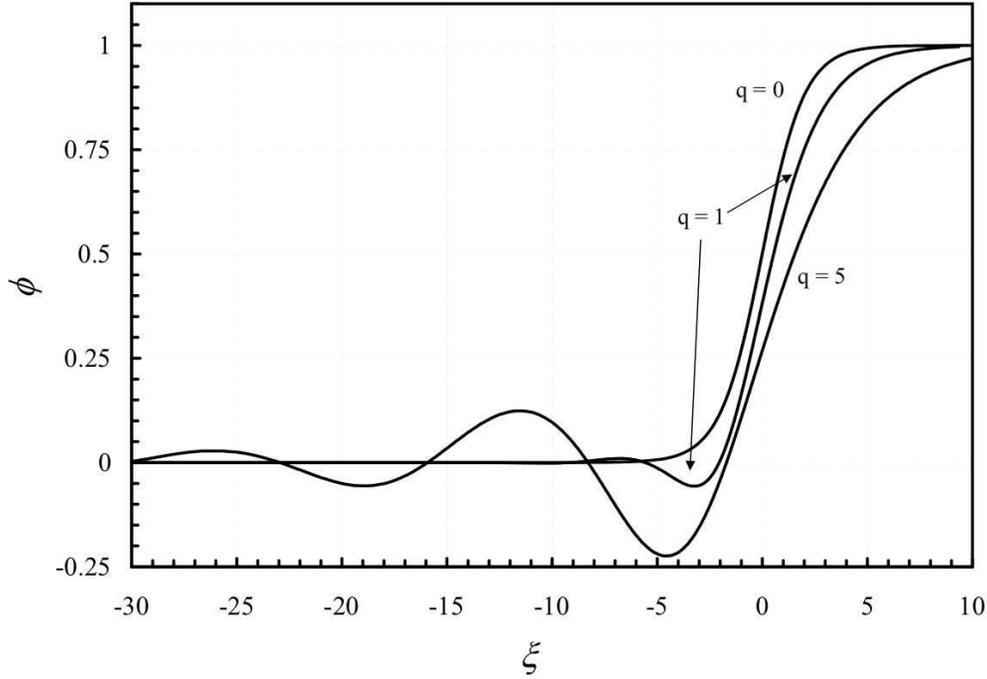}
\caption{Scaled temperature $\phi$
versus scaled coordinate $\xi$ for $q=0, 1, 5$.} \label{fig-2}
\end{figure}

\begin{figure}
\includegraphics[width=.8\columnwidth]{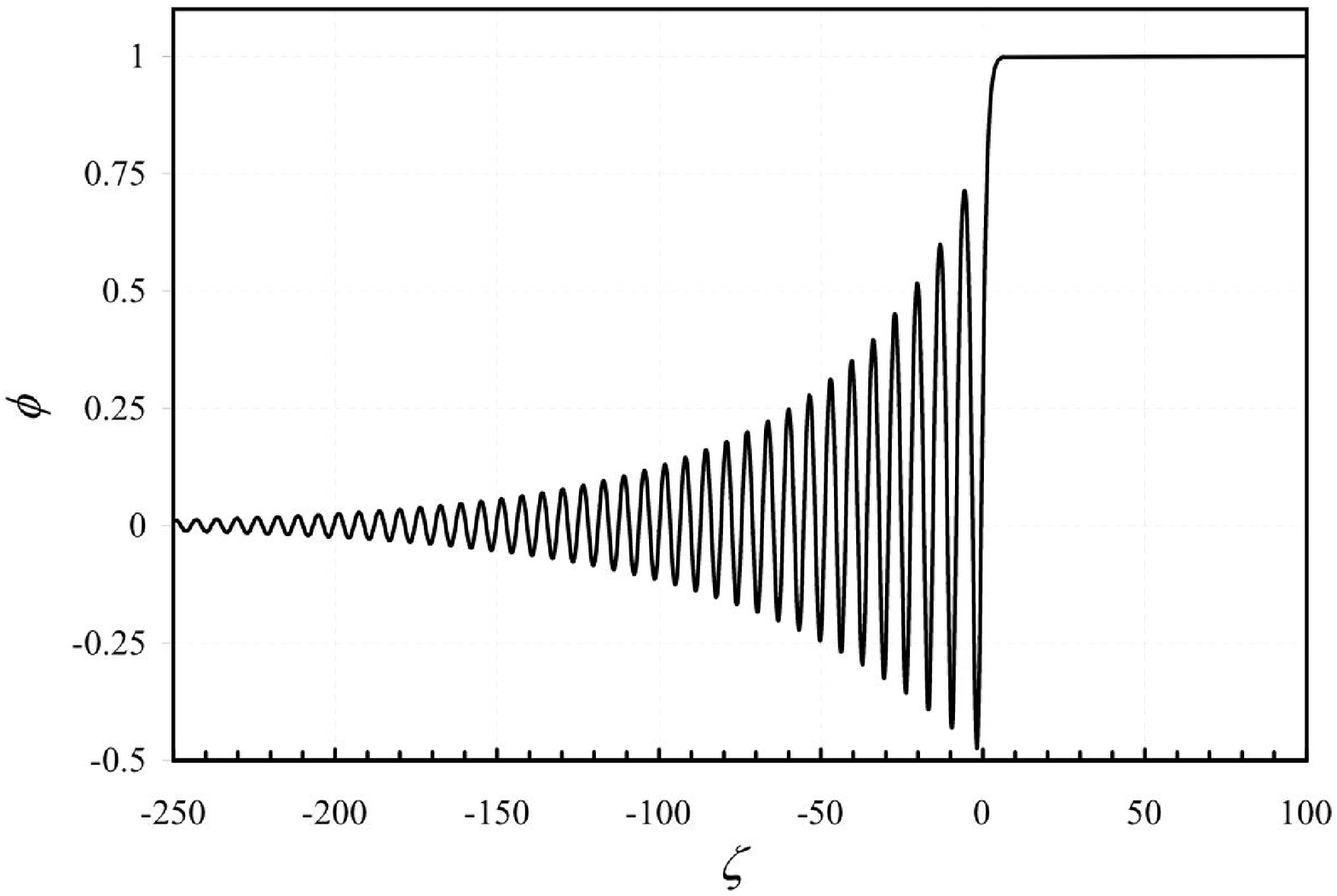}
\caption{Scaled temperature $\phi$
versus scaled coordinate $\zeta = \xi /\sqrt{q}$ for $q=1000$.} \label{fig-3}
\end{figure}

	We can also solve (\ref{eq.29}) analytically in the limit of strong quantum effects,  $q \rightarrow \infty$. In that case the leading terms in (\ref{eq.29}) are
\begin{equation}\label{eq.33}
 \phi(1-\phi)=-q\frac{d^{2}\phi}{d\xi^{2}}.
\end{equation}
Integrating (\ref{eq.33}) once, we obtain
\begin{equation}\label{eq.34}
 \frac{\phi^{2}}{2}-\frac{\phi^{3}}{3}+C=-\frac{q}{2}\left(\frac{d\phi}{d\xi}\right)^{2},
\end{equation}
where $C=-1/6$ due to the boundary condition at the back side of the shock ($\phi \rightarrow 1$). Equation (\ref{eq.34}) may be also rewritten as
\begin{equation}\label{eq.34a}
 (\phi-1)^{2}(2\phi +1)=3q\left(\frac{d\phi}{d\xi}\right)^{2},
\end{equation}
and integrated to give
\begin{equation}\label{eq.35}
  \phi = \frac{3}{2}\tanh^{2}\left(\xi/\sqrt{q}\right)-\frac{1}{2}
  .
\end{equation}
The solution (\ref{eq.35}) is a dark soliton of Korteweg--de Vries type (similar solutions have previously been found in quantum hydrodynamics \cite{brodin-marklund2,marklund-etal}). The solution (\ref{eq.35}) is characterized by a new length scale  $\sqrt{q}$; and it tends to unity, $\phi \rightarrow 1$,  for  $\xi \rightarrow \pm \infty$. We note that neglecting dissipation, we cannot come from initial state $\phi = 0$  to the final state  $\phi = 1$ smoothly. Therefore, the shock inevitably contains a weak discontinuity, which is a surface where  $\phi$ is continuous but $d\phi/d\xi$ has a discontinuity. This weak discontinuity develops at the front side of the shock where $\phi=0$  and
\begin{equation}\label{eq.35a}
\frac{d\phi}{d\xi}=\pm \frac{1}{3\sqrt{q}}.
\end{equation}
The temperature profile may reach the point where $\phi=0$  either from the "bright" or "dark" side, depending on the positive and negative sign, respectively, in Eq. (\ref{eq.35a}). However, in both these cases the transitional region is just a part of the dark soliton solution (\ref{eq.35}), see Fig. \ref{fig-1}. The weak discontinuity may be removed taking into account a small but finite dissipation. In the limit of strong quantum effects and weak dissipation it is more convenient to rescale the space variable as  $\zeta = \xi/\sqrt{q}$. In that case, the dissipation in Eq. (\ref{eq.29}) is small (as $1/\sqrt{q}\ll 1$)
\begin{equation}\label{eq.36}
 \phi(1-\phi)=\frac{1}{\sqrt{q}}\frac{d\phi}{d\zeta}-\frac{d^{2}\phi}{d\zeta^{2}}.
\end{equation}
Dissipation modifies the soliton solution Eq. (\ref{eq.35}) on the front side at  $\zeta \rightarrow - \infty$. Because of the dissipation, $\phi$  cannot reach unity at  $\zeta \rightarrow - \infty$, but tends to zero in the form of decaying oscillations. When  $\phi$ is close to zero, Eq. (\ref{eq.36}) describes small linear oscillations
\begin{equation}\label{eq.37}
 \phi=\frac{1}{\sqrt{q}}\frac{d\phi}{d\zeta}-\frac{d^{2}\phi}{d\zeta^{2}},
\end{equation}
decaying at  $\zeta \rightarrow - \infty$ according to
\begin{equation}\label{eq.38}
 \phi\propto \exp\left[\left(i+\frac{1}{2\sqrt{q}}\right)\zeta\right].
\end{equation}
Equation (\ref{eq.29}) is reduced to a form (\ref{eq.37}) for any non-zero value of $q$ as soon as the temperature comes sufficiently close to the initial value, $\phi \ll 1$, at the front side of the shock. Therefore, even for a small but non-zero quantum effects we should expect oscillations ahead of the shock. The oscillations decay quite fast in the case of relatively small  $q$, but they form a long wave in the limit of large $q$. Numerical solution to Eq. (\ref{eq.29}) is shown in Figs.  \ref{fig-2},  \ref{fig-3} for different values of the parameter  $q= 0; 1; 5; 1000$. The plot with   $q=0$ shows the Burgers solution, which describes a monotonic transition from the initial to the compressed plasma in a classical weak shock. In the case of small but non-zero quantum effects,  $q=1$, we can see one well-pronounced "dark" region with temperature below the initial value ($\phi < 0$). Still, the oscillations decay  fast for $q=1$ and they may be observed only on a small scale of $\phi \ll 1$. Increasing the role of quantum effects,  $q = 5$, we can clearly see a number of peaks and troughs ahead of the shock wave. Finally, in the case of large quantum effects,  $q= 1000$, the front side of the shock looks like a train of oscillations decaying at  $\zeta \rightarrow - \infty$. The last plot resembles a non-stationary train of solitons in a purely quantum medium \cite{Wan-Nature-2007,Kamchatnov.et.al-2002,Kamchatnov.et.al-2004,Damski-2004}. Still, we would like to stress that the shock structure obtained in the present paper is stationary; the train of oscillations propagate together with the shock with the same velocity.


\section{Summary}

In this paper, we have investigated the classical-quantum transition in weak shocks, using a quantum fluid model with finite heat conduction. Both analytical and numerical results were presented, and it was found that soliton trains can occur at the shock front in the quantum regime. Such significant modifications of the front structure could be of interest in laser fusion plasmas.

\acknowledgments

This work has been supported in part by the Swedish Research
Council (VR) and by the Kempe Foundation.

\end{document}